\documentclass[a4paper]{article}
\usepackage{IBERSPEECH2020}

\usepackage[acronym]{glossaries} % load glossaries before hyperref to avoid links
\glsdisablehyper

\usepackage[hyphens]{url}
\usepackage{hyperref}
\hypersetup{breaklinks=true}
\usepackage{cite}
\usepackage{comment}
\usepackage{mathtools}
\usepackage{amsmath,amsfonts,bm}
\usepackage{algorithmic}
\usepackage{graphicx}
\usepackage{array,multirow}
\usepackage{multicol}
\usepackage{comment}
\usepackage{url}
\usepackage[utf8]{inputenc}
\usepackage{mathtools}

\hypersetup{
    colorlinks=true,
    linkcolor=blue,
    filecolor=magenta,      
    urlcolor=blue,
}

\DeclarePairedDelimiter\ceil{\lceil}{\rceil}

\DeclarePairedDelimiterX{\norm}[1]{\lVert}{\rVert}{#1}
\DeclareMathOperator*{\argmax}{argmax} % no space, limits underneath in displays
\DeclareMathOperator*{\argmin}{argmin} % no space, limits underneath in displays
\DeclareMathOperator*{\corr}{corr} % no space, limits underneath in displays
\DeclareMathOperator*{\cov}{cov} % no space, limits underneath in displays
\DeclareMathOperator{\tr}{tr}

\newcommand{\bx}{{\bm x}}
\newcommand{\by}{{\bm y}}
\newcommand{\bX}{{\bm X}}
\newcommand{\bY}{{\bm Y}}
\newcommand{\bZ}{{\bm Z}}
\newcommand{\bz}{{\bm z}}
\newcommand{\bh}{{\bm h}}
\newcommand{\bff}{{\bm f}}
\newcommand{\bg}{{\bm g}}

\newcommand{\bphi}{{\bm \phi}}
\newcommand{\bSigma}{{\bm \Sigma}}

\newacronym{a2a}{A2A}{articulatory-to-acoustic}
\newacronym{asr}{ASR}{automatic speech recognition}
\newacronym{bap}{BAP}{band aperiodicity}
\newacronym{cca}{CCA}{canonical correlation analysis}
\newacronym{ctw}{CTW}{canonical time warping}
\newacronym{dnn}{DNN}{deep neural network}
\newacronym{dtw}{DTW}{dynamic time warping}
\newacronym{dctw}{DCTW}{deep canonical time warping}
\newacronym{emg}{EMG}{electromyography}
\newacronym{kde}{KDE}{kernel density estimation}
\newacronym{gctw}{GCTW}{generalyzed canonical time warping}
\newacronym{lrelu}{LReLU}{leaky rectified linear unit}
\newacronym{mcd}{MCD}{mel-cepstral distortion}
\newacronym{mmi}{MMI}{maximum mutual information}
\newacronym{mse}{MSE}{mean squared error}
\newacronym{mgcc}{MGCC}{mel-generalised cepstral coefficient}
\newacronym{mlpg}{MLPG}{maximum-likelihood parameter generation}
\newacronym{pdf}{pdf}{probability density function}
\newacronym{pca}{PCA}{principal component analysis}
\newacronym{pma}{PMA}{permanent magnet articulography}
\newacronym{relu}{ReLU}{rectified linear unit}
\newacronym{rmse}{RMSE}{root mean squared error}
\newacronym{seq2seq}{seq2seq}{sequence-to-sequence}
\newacronym{ssi}{SSI}{silent speech interface}
\newacronym{kl}{KL}{Kullback-Leibler}
\newacronym{transience}{TRANSIENCE}{multi-view temporal alignment by dependence maximisation in the latent space}

\title{Multi-view Temporal Alignment for Non-parallel Articulatory-to-Acoustic Speech Synthesis}
\name{Jose A. Gonzalez-Lopez$^1$, Miriam Gonzalez-Atienza$^1$, Alejandro Gomez-Alanis$^1$, José L. Pérez-Córdoba$^1$, and Phil D. Green$^2$
\thanks{This work was funded by the Spanish State Research Agency (SRA) under the grant PID2019-108040RB-C22/SRA/10.13039/501100011033. Jose A. Gonzalez-Lopez holds a Juan de la Cierva-Incorporation Fellowship from the Spanish Ministry of Science, Innovation and Universities (IJCI-2017-32926).
}}
\address{
  $^1$University of Granada, Granada, Spain\\
  $^2$University of Sheffield, Sheffield, U.K.}
\email{joseangl@ugr.es}

\begin{document}

\maketitle
\begin{abstract}
\Gls{a2a} synthesis refers to the generation of audible speech from captured movement of the speech articulators. This technique has numerous applications, such as restoring oral communication to people who cannot longer speak due to illness or injury. Most successful techniques so far adopt a supervised learning framework, in which time-synchronous articulatory-and-speech recordings are used to train a supervised machine learning algorithm that can be used later to map articulator movements to speech. This, however, prevents the application of \gls{a2a} techniques in cases where parallel data is unavailable, e.g., a person has already lost her/his voice and only articulatory data can be captured. In this work, we propose a solution to this problem based on the theory of multi-view learning. The proposed algorithm attempts to find an optimal temporal alignment between pairs of non-aligned articulatory-and-acoustic sequences with the same phonetic content by projecting them into a common latent space where both views are maximally correlated and then applying dynamic time warping. Several variants of this idea are discussed and explored. We show that the quality of speech generated in the non-aligned scenario is comparable to that obtained in the parallel scenario.
\end{abstract}

\section{Introduction}
% Loss of speech due to traumatic injury or disease does not only affect the ability of a person to communicate with her/his beloved ones, but also has some important mental health issues associated, such as feelings of personal isolation and social withdrawal, which can lead to clinical depression \cite{Byrne93,Braz2005, Danker2010}. Consequently, different assistive devices have been developed over the years in an attempt to restore  communication capabilities to this sector of society. In this paper, we focus on a particular solution which has attracted increasing attention in recent years: \glspl{ssi} \cite{Denby2010,Gonzalez2020}.

A \gls{ssi} is a type of assistive technology aimed at restoring normal, vocal communication to speech-impaired persons. It does so by decoding speech from biosignals, different from the actual acoustic signal, generated by the human body during speech production. These biosignals can range from the neural activity in the speech and language areas of the brain \cite{Guenther2009,Akbari2019,Anumanchipalli2019}, electrical activity driving the face muscles captured by surface electrodes (i.e., \gls{emg}) \cite{Schultz2010,Wand2014,Janke2017}, or motion capture of the speech articulators by means of imaging techniques \cite{Hueber2010} or electromagnetic articulography techniques \cite{Schonle87,Fagan2008,Gonzalez2016,Gonzalez2017}. 

Because \glspl{ssi} do not rely on the acoustic speech signal, they offer a radically new form of restoring oral communication to people with speech impairments. Two alternative \gls{ssi} approaches have been proposed to decode speech from speech-related biosignals \cite{Gonzalez2020}: silent speech recognition, which involves the use of \gls{asr} algorithms to transform the biosignals into text, and direct speech synthesis, which directly maps the biosignals into a set of acoustic parameters amenable to speech synthesis. In this work, we focus on the latter approach, which is also known as \glsreset{a2a} \gls{a2a} synthesis in the literature when the biosignals encode information about the movements of the speech organs. 

Most successful direct synthesis techniques so far adopt a data-driven framework in which supervised machine learning is used to model the mapping $\by = \bh(\bx)$ between source feature vectors $\bx$ extracted from the biosignals and target feature vectors $\by$ computed from the speech signals. To train this function, a dataset $\mathcal{D}=\lbrace (\bX_1, \bY_1),\ldots, (\bX_M, \bY_M)\rbrace$ with pairs of sequences of feature vectors $(\bX_i, \bY_i)$ extracted from time-synchronous recordings is used, where $\bX_i \in \mathbb{R}^{d_x \times T_i}$ and $\bY_i \in \mathbb{R}^{d_y \times T_i}$. The need for parallel recordings, however, limits the application of direct synthesis techniques to only a few clinical scenarios, as described in \cite{Gonzalez2020}. For instance, people who have already lost their voices could not use this technology because of the impossibility of recording parallel data.

A solution to this problem, which we explore in this work, involves the use of speech recordings from voice donors (e.g., relatives or recordings of the patient's own voice made before the voice loss). Using these recordings, it would be possible in principle to obtain the necessary parallel data by asking the patient to mouth along the speech recordings while articulatory data is captured. Even in this case, it is likely that the articulatory data will not be perfectly aligned with the speech recordings, having both modalities slightly different duration, thus preventing the direct application of standard supervised machine learning techniques.

To address this issue, in this work we propose an algorithm called \gls{transience}\footnote{Code is available at \url{https://github.com/joseangl/transience}.}, which is based on the theory of multi-view learning \cite{Andrew2013,WWang2015}. \Gls{transience} attempts to find the optimal temporal alignment between sequences of multiple views (e.g., audio and articulatory data) by first non-linearly projecting the data into a common, latent subspace where the views are maximally dependent and then aligning the resulting latent variables by means of the \gls{dtw} algorithm\footnote{It should be noted that the direct application of \gls{dtw} to this problem is not possible because the views may have different dimensionality.} \cite{Rabiner93}. We examine the performance of the proposed algorithm on a \gls{a2a} task involving the conversion of articulatory data captured using \gls{pma}  \cite{Fang2018,Gilbert2010} to speech for multiple non-impaired subjects.

The remainder of this paper is organised as follows. First, in Section \ref{sec:related-work}, the relevant related work is reviewed and the main differences w.r.t. our technique are discussed. The details of the proposed technique are presented in Section \ref{sec:algorithm}. Section \ref{sec:exp-setup} describes the experimental setup, while the results are shown in Section \ref{sec:results}. We conclude in Section \ref{sec:conclusions} and discuss potential future research directions.

\section{Related work}
\label{sec:related-work}
The most closely related work to our method is that of Trigeorgis \emph{et al.} \cite{Trigeorgis2017}, in which an algorithm called \gls{dctw} combining \gls{cca} \cite{Hotelling1936} with \gls{dtw} is proposed. The key differences of our approach w.r.t. \gls{dctw} can be summarised as follows. First, we evaluate different similarity metrics, not only \gls{cca}, to optimise the parameters of the \glspl{dnn} used to map the multiple views into their common, latent subspace. Also, we introduce an autoenconder-based loss, which helps to regularise the training and avoids \emph{naïve} solutions. Finally, inspired by the work in \cite{WWang2016}, we also propose the introduction of private latent variables for each view which aim at modelling the specific peculiarities within each view. Our technique also shares similarities with the \gls{gctw} technique described in \cite{Zhou2015}. However, in contrast to this technique, \gls{transience} solves the optimal alignment problem with \gls{dtw} rather than approximating the temporal warping with a set of pre-defined monotonic bases and optimising the weights of these bases with a Gauss-Newton algorithm. Also, being based on \gls{cca}, \gls{gctw} computes the latent variables by \emph{linearly projecting} the data from the different views, while our method uses powerful autoencoders to non-linearly transform the data, which we could be expected to yield better alignments.

Contrary to now popular \gls{seq2seq} models (e.g., \cite{Sutskever2014,Vaswani2017}), our method has different advantages. First, our aim is to align sequences of different lengths in order to obtain the necessary parallel data to train a machine learning technique, not to model a full-fledged mapping between sequences. In other words, our technique is only used to align the training data, while in test time the articulatory data is directly mapped to speech. Also, it is relatively straightforward to modify our method to process multiple views (more than 2), while the adaptation of \gls{seq2seq} models is more involved. 

\section{Multi-view temporal alignment}
\label{sec:algorithm}
The problem we address can be formulated as finding the optimal alignment between two time series $\bX= (\bx_1, \ldots, \bx_{T_x})$ and $\bY= (\by_1, \ldots, \by_{T_y})$ with possibly different dimensions. In our case, $\bX \in \mathbb{R}^{d_x \times T_x}$ is a sequence of feature vectors extracted from an articulatory signal and $\bY \in \mathbb{R}^{d_y \times T_y}$ is the sequence of acoustic speech parameters. We further assume that both sequences encode the \emph{same phonetic content} (i.e., same words in the same order) but have, possibly, different duration. Mathematically, this involves solving the following minimisation problem,
\begin{equation}
    \argmin_{\bphi^x, \bphi^y} \sum_{t=1}^T d(\bx_{\phi_t^x},\by_{\phi_t^y}),
    \label{equ:problem1}
\end{equation}

\noindent where $T= \max(T_x, T_y)$, $d(\bx,\by)$ is a distance function, $\bphi^x \in \lbrace1:T_x\rbrace^T$ and $\bphi^y \in \lbrace1:T_y\rbrace^T$ are warping functions that map the indices of the original time series to a common time axis. This way, $\bx_i$ and $\by_j$ will be aligned if $\phi_t^x=i$ and $\phi_t^y=j$ for a given $t$.

To enable the alignment of sequences with different dimensionality, we assume that there exists a pair of transformation functions $\bff: \mathbb{R}^{d_x} \to \mathbb{R}^{d_z}$ and $\bg: \mathbb{R}^{d_y} \to \mathbb{R}^{d_z}$, modelled as \glspl{dnn} in this work, that map the feature vectors into a common, latent space where the data from both views are maximally dependent (in the next section we will discuss how we measure this dependence). Thus, the problem in \eqref{equ:problem1} now becomes,  
\begin{equation}
    \argmin_{\bphi^x, \bphi^y}  \sum_{t=1}^T d(\bff(\bx_{\phi_t^x}),\bg(\by_{\phi_t^y})).
    \label{equ:problem2}
\end{equation}

For some fixed mapping functions $\bz^x = \bff(\bx)$ and $\bz^y = \bg(\by)$, the problem of temporal alignment of latent variable sequences $\bZ^x = (\bz_1^x, \ldots, \bz_{T_x}^x)$ and $\bZ^y = (\bz_1^y, \ldots, \bz_{T_y}^y)$ can be solved efficiently by means of the \gls{dtw} algorithm \cite{Rabiner93}. Conversely, for fixed warping paths $\bphi^x$ and $\bphi^y$, the problem in \eqref{equ:problem2} involves optimising the functions $\bff(\cdot)$ and $\bg(\cdot)$ in order to minimise $\sum_{t=1}^T d(\bff(\bx_{\phi_t^x}),\bg(\by_{\phi_t^y}))$. If the mapping functions are modelled as \glspl{dnn}, as in our case, this latter problem can be solved by back-propagation. Thus, \gls{transience} algorithm solves \eqref{equ:problem2} by alternating between the following two phases: (i) finding the optimum \glspl{dnn} weights $(\Theta_f, \Theta_g)$ by fixing the warping paths, and (2) applying \gls{dtw} to compute the optimum warping paths (i.e., optimum alignments) $(\phi^x, \phi^y)$ by freezing the \glspl{dnn} weights. The warping paths are initialised by uniformly aligning the sequences, i.e., $\phi_t^x= 1+ \ceil*{\frac{t-1}{T-1}(T_x-1)}$ and $\phi_t^y= 1+ \ceil*{\frac{t-1}{T-1}(T_y-1)}$ for $t=1, \ldots, T$ and $T = \max(T_x,T_y)$. 

% A block diagram of the proposed algorithm is shown in Fig. \ref{fig:block_diagram}.

% \begin{figure*}[t]
% \centering	
% \includegraphics[width=0.7\textwidth]{figs/block_diagram.pdf}
% \caption{Block diagram of the proposed \gls{transience} algorithm for multi-view temporal alignment. Figure adapted from \cite{Trigeorgis2017}.}
% \label{fig:block_diagram}
% \end{figure*}

\subsection{Latent-space dependence metrics}
\label{ssec:sim-metrics}
A key component of our algorithm is the distance function $d(\bff(\bx), \bg(\by))$ used in \eqref{equ:problem2} to evaluate the dependence between pairs of aligned latent variables. Here, we evaluate three alternative loss functions that optimise this dependence by maximising the correlation, mutual information and (minimise) a contrastive loss, respectively, between the latent variables.

\subsubsection{Canonical correlation analysis}
\label{sssec:cca}
Given a mini-batch of $N$ pairs of \emph{aligned} observations $\mathcal{B}= \lbrace (\bx_1,\by_1), \ldots (\bx_N,\by_N) \rbrace$, this loss function attempts to maximise the correlation between the outputs of the \glspl{dnn}, $\bff(\bx)$ and $\bg(\by)$, as follows,
\begin{equation}
    (\Theta_f^*,\Theta_g^*) = \argmax_{\Theta_f,\Theta_g} \sum_{i=1}^N\corr(\bff(\bx_i; \Theta_f),\bg(\by_i;\Theta_g)).
\end{equation}

As detailed in \cite{Andrew2013}, this equals to maximising the following loss function,
\begin{equation}
    \mathcal{L}_{cca} = \sqrt{\tr\left(\bm{T}' \bm{T}\right)},
    \label{equ:cca-loss}
\end{equation}

\noindent where $\tr(\cdot)$ is the trace operator and $\bm{T} = \bSigma_{xx}^{-1/2} \bSigma_{xy} \bSigma_{yy}^{-1/2}$. The covariance matrices $\bSigma_{xx} = \cov(\bff(\bx), \bff(\bx))$, $\bSigma_{xy} = \cov(\bff(\bx), \bg(\by))$ and $\bSigma_{yy} = \cov(\bg(\by), \bg(\by))$ are estimated from the outputs of the \glspl{dnn}. It should be noted that the \gls{dctw} algorithm described in \cite{Trigeorgis2017} is a specific case of our \gls{transience} algorithm when the \gls{cca} loss in \eqref{equ:cca-loss} is used.

\subsubsection{Maximum mutual information}
\label{sssec:mmi}
As an alternative, we also consider maximising the mutual information between the latent variables as follows,
\begin{equation}
    \mathcal{L}_{mmi} = \sum_{i=1}^N p(\bff(\bx_i),\bg(\by_i)) \log \frac{p(\bff(\bx_i),\bg(\by_i))}{p(\bff(\bx_i)) p(\bg(\by_i))} .
    \label{equ:mmi-loss}
\end{equation}

The \glspl{pdf} in \eqref{equ:mmi-loss} are estimated using \gls{kde} as follows,

\begin{equation}
    p(\bz_i) = \frac{1}{N-1} \sum_{j=1, j \neq i}^N K(\bz_i - \bz_j),
    \label{equ:kde}
\end{equation}

\noindent where an isotropic Gaussian kernel with trainable bandwidth $\sigma_z$ is used in this work, i.e., $K(\bz) = \mathcal{N}(\bz; \bm{0}, \sigma_z \bm{I})$. Thus, three \glspl{pdf} are estimated, the joint distribution $p(\bff(\bx),\bg(\by))$ and the marginals $p(\bff(\bx))$ and $p(\bg(\by))$, each one with its own trainable bandwidth.
 
\subsubsection{Contrastive loss}
\label{sssec:contrastive}
Finally, we also evaluate the contrastive loss function described in \cite{Hermann2014,WWang2016} which, given a fixed latent variable from the first view $\bff(\bx^+)$, takes an aligned positive example $\bg(\by^+)$ and an unaligned negative example $\bg(\by^-)$ from the second view and attempts to \emph{minimise} the difference between the distances for the positive and negative examples:
\begin{align}
    \mathcal{L}_{contrastive} = \frac{1}{N} \sum_{i=1}^N & \max(0, m +  d(\bff(\bx_i^+),\bg(\by_i^+)) \nonumber \\
    & - d(\bff(\bx_i^+),\bg(\by_i^-))),
    \label{equ:contrastive-loss}
\end{align}

\noindent where $m$ is a margin hyperparameter ($m=0.5$ is used in this work) and $d(\bz_x, \bz_y)= 1 - \frac{\bz_x \cdot \bz_y}{\norm{\bz_x} \norm{\bz_y}}$ is the cosine similarity. The negative examples $\bg(\by^-)$ are generated by shuffling the outputs of the \gls{dnn} for the second view before the loss is computed. Intuitively, the distances $d(\bff(\bx^+),\bg(\by^+))$ in \eqref{equ:contrastive-loss} should be small if both views are projected to similar (closer) representations in the common latent space, whereas the distances to the negative, unpaired examples $d(\bff(\bx^+),\bg(\by^-))$ should be bigger because they are projected to different locations of that space.

\subsection{Multi-view autoencoder}
\label{ssec:autoencoder}
In order to regularise the training of the \glspl{dnn} and to avoid naïve solutions (e.g., $\bff(\bx) = \bg(\by) = \bm{c}$, for all $\bx$ and $\by$, with $\bm{c}$ being a constant vector), we introduce an autoencoder-based reconstruction loss that minimises the \gls{mse} between the \glspl{dnn}' inputs $(\bx, \by)$ and the reconstructed outputs $(\hat{\bx}=\bff^{-1}(\bff(\bx)), \hat{\by}= \bg^{-1}(\bg(\by)))$, being $\bff^{-1}$ and $\bg^{-1}$ decoder networks that attempt to reconstruct $\bx$ and $\by$ from their latent projections. The parameters of such networks, as well as those from the encoders $\bff$ and $\bg$, are trained by gradient-descend techniques using the following loss function,
\begin{align}
    \mathcal{L}_{autoencoder} &= \frac{\lambda}{N} \left( \sum_{i=1}^N \norm[\Big]{\bx_i - \bff^{-1}(\bff(\bx_i))}^2 \right. \nonumber \\
    & \qquad + \left. \sum_{i=1}^n \norm[\Big]{\by_i - \bg^{-1}(\bg(\by_i))}^2 \right),
    \label{equ:reconstruction-loss}
\end{align}
\noindent where the hyperparameter $\lambda$ is set to $1$ in this work.

\subsection{Private latent variables}
\label{ssec:latent-variables}
A key assumption of \gls{transience} is that the views share some common information. Although this is indeed the case for audio and articulatory data, where both views encode the same phonetic information, it is also true that each view may have its own unique characteristics, thus making the reconstruction loss in \eqref{equ:reconstruction-loss} difficult to optimise when only considering the shared latent variables. For instance, the \gls{pma} technique used in this work for articulator motion capture is known to model poorly the information about the speech prosody \cite{Gonzalez2014,Gonzalez2017b}, whereas this information can be easily decoded from the acoustic signal. Thus, it may be beneficial to model the unique characteristics of each view as well as the common characteristics shared among all the views. Inspired by the work in \cite{WWang2016}, we propose the introduction of private latent variables for each view $\tilde{\bz}^x$ and $\tilde{\bz}^y$ that aim at modelling the uniqueness of each view. The private variables are predicted from the inputs by a set of independent \glspl{dnn} $\tilde{\bz}^x = \tilde{\bff}(\bx)$ and $\tilde{\bz}^y = \tilde{\bg}(\by)$. These private latent variables are used in \eqref{equ:reconstruction-loss}, in addition to the common variables, for reconstructing the input data, i.e., $\hat{\bx} = \bff^{-1}(\bff(\bx), \tilde{\bff}(\bx))$ and $\hat{\by} = \bg^{-1}(\bg(\by), \tilde{\bg}(\by))$.

When optimising the weights of \glspl{dnn} $\tilde{\bff}$ and $\tilde{\bg}$, a standard Gaussian distribution $\tilde{\bz} \sim \mathcal{N}(\bm{0}, \bm{I})$ is chosen as the prior distribution for the private variables of each view. To enforce this distribution, we minimise the \gls{kl} divergence between the priors and the empirical distribution (modelled as a multivariate Gaussian distribution with diagonal covariance) estimated from the private variables as follows,
\begin{equation}
    \mathcal{L}_{KL} = \frac{1}{2} \sum_{i=1}^{d_{\tilde{z}}} \left(\sigma_i^2 + \mu_i^2 - 1 - \log\sigma_i^2 \right),
    \label{equ:kl-loss}
\end{equation}
\noindent where the mean and variances in \eqref{equ:kl-loss} are estimated for each private variable in each mini-batch.

\section{Experimental setup}
\label{sec:exp-setup}
\subsection{Dataset}
The proposed alignment algorithm was evaluated on a \gls{a2a} task involving the synthesis of speech from articulatory data captured using \gls{pma} \cite{Fagan2008,Gilbert2010}. Parallel data was recorded by four non-impaired British subjects (2 males and 2 females) while reading aloud a subset of the CMU Arctic corpus \cite{Kominek2004}. Two alignment conditions were evaluated: (i) intra-subject alignment, where \gls{pma} and speech signals recorded by the same subject in different sessions are aligned, and (ii) cross-subject alignment, where \gls{pma} signals recorded by a given subject are aligned with speech recorded by a different subject (with possibly different gender as well). We also attempted to align \gls{pma} signals recorded from a female laryngectomy patient with a set of speech recordings made by her before losing the voice. The details of the dataset used for our experiments is summarised in Table \ref{tab:dataset}.

\begin{table}[t]
\caption{\label{tab:dataset} Details of the dataset used for the experiments.}
\vspace*{-0.25cm}
\centering
	\begin{footnotesize}
	\begin{tabular}{ l l  c  c  }
    \hline
    &&  \multicolumn{2}{c}{\textbf{\# of sentences}} \\
    \multicolumn{2}{c}{\textbf{Condition}} & \emph{Train} & \emph{Eval.} \\
    \hline
    Intra-subject   & F $\rightarrow$ F & 103 (11.3 min) & 20  (1.2 min) \\
                    & M $\rightarrow$ M & 134 (8.4 min) & 20 (1.1 min) \\
    \hline     
    Cross-subject   & F $\rightarrow$ F & 99 (6.0 min) & 18 (0.9 min) \\
                    & F $\rightarrow$ M & 332 (18.7 min) & 20 (1.0 min) \\
                    & M $\rightarrow$ F & 332 (21.9 min) & 20 (1.2 min)\\
                    & M $\rightarrow$ M & 414 (27.9 min) & 20 (1.2 min)\\
    \hline       
    \end{tabular}
    	\end{footnotesize}
\vspace*{-0.5cm}
\end{table}

\subsection{Implementation details}
Each \gls{dnn} in \gls{transience} was modelled as a 3-layer feed-forward neural network with $200 \times 100 \times 100$ hidden units and \gls{lrelu} activations ($a= 0.03$) following \cite{Trigeorgis2017}. The neural networks were trained as denoising autoencoders ($\sigma_{noise} = 0.5$) using the Adam algorithm \cite{Kingma2014} with a fixed learning rate of $1e-4$ and a batch size of $N= 512$ samples. The dimensionality of the shared latent variables was set to $d_z = 20$ and fixed $d_{\tilde{z}}= 10$ for the private variables. The \gls{pma} signals were parameterised by applying \gls{pca} over contextual windows with 11 frames. The acoustic signals, on the other hand, were parameterised using the WORLD vocoder \cite{Morise2016} with a frame rate of 5 ms as 25 \glspl{mgcc}, 1 \gls{bap} value, 1 continuous $F_0$ value in logarithmic scale and 1 U/V decision. For temporal alignment, only the \glspl{mgcc} were used (augmented with delta and acceleration parameters). Finally, the cosine distance was used for \gls{dtw}.

\subsection{PMA-to-speech system}
The aligned signals were used to train speaker-dependent \gls{a2a} systems. We used the same setting as in our previous work \cite{Gonzalez2017b}: \glspl{dnn} with 4 hidden layers with 400 units in each layer and \gls{relu} activations were used. The \gls{mlpg} algorithm \cite{Tokuda2000, Toda2007} was applied over the \gls{dnn} outputs to enhance the acoustic quality of the re-synthesised waveforms.

\section{Results}
\label{sec:results}
\subsection{Objective evaluation}
\begin{table}[t]
\caption{\label{tab:results} Summary of the objective results.}
\vspace*{-0.25cm}
	\setlength{\tabcolsep}{4pt}
\centering
	\begin{footnotesize}
	\begin{tabular}{ l  c  c c c  }
     \hline
       & \textbf{MGCC} & \textbf{BAP} & $\bf{F_0}$ & \textbf{Voicing}\\
    \textbf{Method} & MCD (dB) & RMSE (dB) & RMSE (Hz) & err. rate(\%) \\
    \hline
    Oracle & 7.81 & 0.43 & 14.75 & 23.79\\
    \hline
    CTW & 8.55 & 0.59 & 15.98 & 23.08 \\
    +autoenc. & 9.20 & 0.88 & 16.70 & 25.30 \\
    +priv. vars. & 8.83 & 0.58 & 15.74 & \textbf{21.47} \\
    \hline
    CCA & 9.37 & 0.85  & 16.40 & 27.95 \\
    +autoenc. & 10.02 & 1.46 & 15.95 & 34.08 \\
    +priv. vars. & 10.48 & 1.24 & 15.79 & 31.88 \\
    \hline
    MMI & 9.74 & 0.69 & 16.43 & 22.25 \\
    +autoenc. & 9.97&  1.09 & 16.92  & 23.72\\
    +priv. vars. & 9.86 & 1.70 & 16.41 & 21.75 \\
    \hline
    Contrastive & \textbf{7.65} & \textbf{0.12} & 15.28 & 24.10 \\
    +autoenc. & 7.76 & 0.20& 14.98 & 24.06 \\
    +priv. vars. & 7.82& 0.30& \textbf{14.58} & 23.68 \\    
    \hline       
    \end{tabular}
    	\end{footnotesize}
\vspace*{-0.6cm}
\end{table}

First, we evaluated the quality of the re-synthesised speech signals obtained from the test \gls{pma} signals by comparing them with the original speech recordings made by the non-impaired subjects. For this task, we used several objective metrics widely used in speech synthesis: \gls{mcd}, \gls{rmse} of the predicted \gls{bap} and $F_0$ values and the error rate for the voicing parameter. For \gls{transience}, three variants were evaluated depending on the latent-space similarity loss function employed: \gls{cca}, \gls{mmi}, and Contrastive losses described in Section \ref{ssec:sim-metrics}. Furthermore, for each of the three variants, we also evaluated the effect of the autoencoder-based loss described in Section \ref{ssec:autoencoder} and the introduction of the private variables described in Section \ref{ssec:latent-variables}. For comparison purposes, we also evaluated the \gls{ctw} technique described in \cite{Zhou2009}, which is a particular case of \gls{transience} combining standard (linear) \gls{cca} with \gls{dtw}, thus not being able to model non-linear latent mappings. We also provide the results obtained by an oracle system, in which the \gls{pma} and acoustic signals in the training dataset are aligned by using the \emph{ideal} warping paths computed by applying \gls{dtw} over the speech signals recorded by the same subjects.

Table \ref{tab:results} shows the objective results for the different systems. \gls{transience} using the contrastive loss yields the best alignments and, hence, the best objective results, outperforming the other similarity metrics and, even, the oracle system. In particular, relative gains of 18.36\% and 21.46\% are achieved in the \gls{mcd} metric w.r.t. using the \gls{cca} and \gls{mmi} losses. Unfortunately, it seems that the introduction of the autoencoder-based loss and the private latent variables do not improve the results, which may be due to the dimensionality of the private variables not being enough to capture the peculiarities of each view. It is also surprising that the simple (linear) \gls{ctw} technique outperforms its non-linear version \gls{transience}+\gls{cca}. In future work, we should look at this issue.

\subsection{Subjective evaluation}
\begin{figure}[t]
\centering	
\includegraphics[scale=0.75]{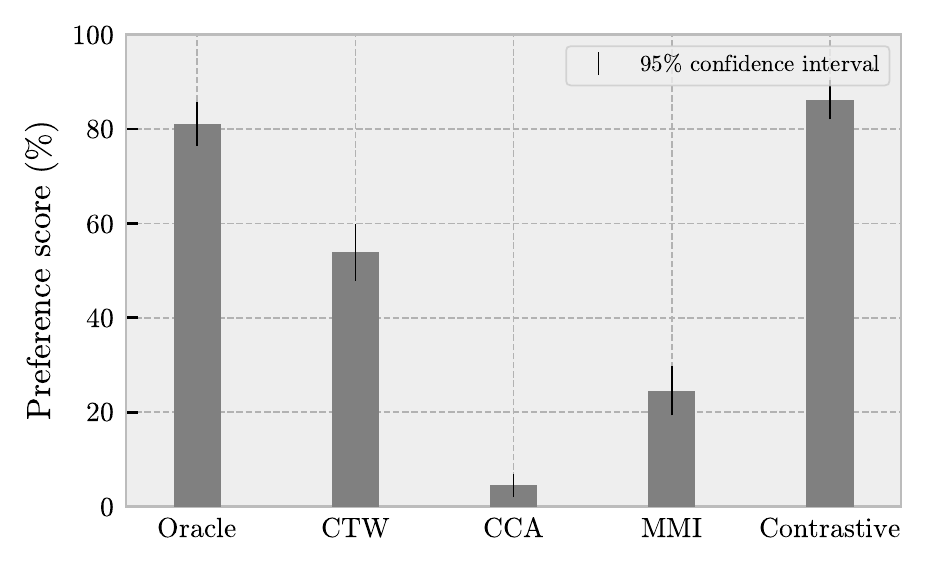}
\vspace*{-0.3cm}
\caption{\label{fig:xab} Results of the ABX test on speech quality.}
\vspace*{-0.6cm}
\end{figure}

We also conducted an ABX test to subjectively evaluate the quality of the resynthesised speech signals. 27 listeners participated in the test, who have to judge which of two versions of the same signal produced by any combination of two of the 5 systems in Table \ref{tab:results} was more similar to a reference (one of the signals recorded by the subjects). Each listeners evaluated 10 sample pairs for each of the 10 possible system combinations (i.e., 100 pairs evaluated in total). For this task, only the ''basic´´ systems in Table \ref{tab:results} were evaluated (i.e., without autoencoder-loss and private variables), because this setting produced the best objective results. Fig. \ref{fig:xab} shows the results of the listening test. The most preferred system by a large margin was Contrastive, being on par with the Oracle system. Interestingly, the \gls{ctw} system obtained higher preference scores than its non-linear version (\gls{transience}+\gls{cca}) and that the \gls{mmi} system. It may be due that the optimisation process get stuck in poor local-minima for the latter systems. However, more research is needed to shed some light into this problem.

\section{Conclusions}
\label{sec:conclusions}
We have proposed a new method for the alignment of times series with different dimensionality. Our evaluation on a \gls{a2a} task involving the synthesis of audible speech from articulatory signals has shown that it is feasible to deploy direct synthesis techniques in non-parallel scenarios. Future work include evaluating our technique using more data from clinical population and introducing more constraints for the alignment (e.g., phonetic constraints).

%\section{Acknowledgements}
% This work was funded by the Spanish State Research Agency (SRA) under the grant PID2019-108040RB-C22/SRA/10.13039/501100011033. Jose A. Gonzalez-Lopez holds a Juan de la Cierva-Incorporation Fellowship from the Spanish Ministry of Science, Innovation and Universities (IJCI-2017-32926).

\bibliographystyle{IEEEtran}
\bibliography{journal_abbreviations,paper}

\end{document}